# Are My EHRs Private Enough?
# - Event-level Privacy Protection

Chengsheng Mao*, Yuan Zhao*, Mengxin Sun, Yuan Luo

**Abstract**— Privacy is a major concern in sharing human subject data to researchers for secondary analyses. A simple binary consent (opt-in or not) may significantly reduce the amount of sharable data, since many patients might only be concerned about a few sensitive medical conditions rather than the entire medical records. We propose event-level privacy protection, and develop a feature ablation method to protect event-level privacy in electronic medical records. Using a list of 13 sensitive diagnoses, we evaluate the feasibility and the efficacy of the proposed method. As feature ablation progresses, the identifiability of a sensitive medical condition decreases with varying speeds on different diseases. We find that these sensitive diagnoses can be divided into 3 categories: (1) 5 diseases have fast declining identifiability (AUC below 0.6 with less than 400 features excluded); (2) 7 diseases with progressively declining identifiability (AUC below 0.7 with between 200 and 700 features excluded); and (3) 1 disease with slowly declining identifiability (AUC above 0.7 with 1000 features excluded). The fact that the majority (12 out of 13) of the sensitive diseases fall into the first two categories suggests the potential of the proposed feature ablation method as a solution for event-level privacy protection.

—————————— ◆ ——————————

## 1 INTRODUCTION

Data privacy is a major concern in biomedical data sharing and analysis. Electronic healthcare records (EHRs) contain sensitive information that can risk the privacy of individuals, which therefore needs to be carefully processed. Despite the policies on protecting data privacy, research has demonstrated that the privacy risk of re-identification still exists even the data are HIPAA (Health Insurance Portability and Accountability Act) [1] de-identified [2-4]. In order to keep sensitive patient information confidential, we need to develop technical solutions to safeguard patients' privacy in EHRs. This is a challenging task and we need to consider the trade-off between privacy and utility. We may consider adopting strong privacy-preserving solutions and publish only a few data points for research, which might hurt the validity of the research outcomes. On the other hand, if we do not sufficiently suppress sensitive information, patient identity can be at risk [5, 6]. This is often a dilemma to the biomedical and privacy research community.

There is a sweet spot, however, if we carefully investigate this problem, which might allow us to develop novel solutions to strike the right balance between privacy and utility. That is, in reality, patients do not view all their data being equally sensitive because most of the time only a small proportion might be deemed confidential as reported in several studies that have looked into personalized protection of sensitive information [7, 8]. We are inspired by this idea and believe it is possible to separate sensitive information from the medical records of individual patients in a fine-grained manner (e.g., a patient may be reluctant to reveal his/her depression but willing to disclose other conditions) to support event-level privacy protection. With the ability to remove sensitive diseases and their associated conditions, patients can have more options regarding how they want to share their data and might encourage data sharing with reduced privacy concerns. This seemingly simple idea is indeed non-trivial because suppression of explicitly sensitive information (e.g., diagnosis) is not sufficient to protect patient privacy. For example, patients with depressive disorder (ICD-9-CM Diagnosis Code 311) might take anti-depressants like citalopram, escitalopram, and paroxetine, among other medications. The medication events could lead to the identification of the disease event even the diagnosis code 311 is removed.

In data privacy, many state-of-the-art approaches adopt differential privacy [9], a strong crypto-motivated criterion, to mask sensitive information by noise perturbation. Under this principle, researchers have developed parametric, non-parametric, and semiparametric models [10-13] to protect information dissemination. A recent survey summarizes a number of methods and their biomedical applications [14]. However, there are still limitations with these approaches. First, the noise added to the data might introduce incorrect information that could lead to medical errors. Second, perturbing the entire database (e.g., sufficient statistics) might be useful for hypothesis testing and

________________

- *Chengsheng Mao is with the Department of Preventive Medicine, Northwestern University, Chicago, IL 60611. E-mail: chengsheng.mao@northwestern.edu.*
- *Yuan Zhao is with the Department of Preventive Medicine, Northwestern University, Chicago, IL 60611. E-mail: yuan.zhao1@northwestern.edu.*
- *Mengxin (Ivy) Sun is with the Northwestern Medicine, Chicago, IL 60611. E-mail: mengxin.sun@nm.org.*
- *Yuan Luo is with the Department of Preventive Medicine, Northwestern University, Chicago, IL 60611. E-mail: yuan.luo@northwestern.edu.*

machine learning but individual trajectories of the medical records might be destroyed. A recent study proposes the concept of "personalized privacy" [15], which allows individuals to set their own thresholds to determine the degree of protection. However, these methods do not consider the "event-level privacy" of individual patients as the methods do not discriminate different events of individual patients' EHRs.

The notion of event-level privacy has an interpretation from a novel perspective of computational phenotyping. Computational phenotyping is a method that aims to automatically mine or predict clinically significant (or scientifically meaningful) phenotypes from structured EHR data, unstructured clinical narratives, or their combinations. In medicine, the term phenotype refers to observable properties of a cohort of patients based on the interactions of their genotypes and the environment [16]. The goal of computational phenotyping is to extract a phenotype from complex and heterogeneous data sources and/or to predict clinically important phenotypes before they are actually observed. Depending on the nature of the task, computational phenotyping can be formulated as either an unsupervised or supervised learning problem. As summarized in the review by Shivade *et al.* [17], supervised computational phenotyping studies are provided with predefined phenotypes, and the task is to identify a patient cohort matching the definition's criteria. Many of these studies relied heavily on structured and coded patient data, where structured data typically capture patients' events of laboratory test results, medication prescriptions, and procedure codes [18]. Event-level privacy protection can be viewed as a supervised computational phenotyping task where the sensitive disease event is the phenotype (e.g., ICD9 code 311). To ensure event-level privacy, our task is to obliterate events in the data that would allow the identification of sensitive disease events. Conventional computational phenotyping strives for high performance (e.g. Area Under the ROC Curve), while event-level privacy protection favors low identifiability. The models are further complicated by heterogeneous data sources such as those from multiple instituions [17]. To take the data heterogeneity factor into consideration, in this study, we deliberately collect our patient EHRs from two hospitals.

## 2 METHODS

### 2.1 Sensitive disease list

We selected 13 sensitive diseases as shown in Table 1. The list of sensitive diseases was categorized into 6 groups, including Sexually transmitted diseases, Mental Diseases, Drug Addiction/Abuse, Male Reproductive Related, Female Reproductive Related, and Diseases of Newborn. This list includes typical diseases of each categories. Though not comprehensive, we believe that the list serves a proof-of-concept to understand the sensitivity of different diseases that are common in EHRs.

Table 1 Sensitive disease ICD9 codes, cohort sizes, descriptions, and the categories.

| ICD9 | Number of Patients | Description | Category |
| --- | --- | --- | --- |
| 042 | 2462 | Human immunodeficiency virus (HIV) | Sexually transmitted diseases |
| 099 | 945 | Other venereal diseases | |
| 300 | 5000 | Anxiety, dissociative and somatoform disorders | Mental Diseases |
| 311 | 5000 | Depressive disorder, not elsewhere classified | |
| 304 | 2540 | Drug dependence | Drug Addiction/Abuse |
| 305 | 5000 | Nondependent abuse of drugs | |
| 306 | 1645 | Physiological malfunction arising from mental factors | |
| 606 | 3142 | Infertility, male | Male reproductive related |
| 607 | 5000 | Disorders of penis | |
| 626 | 5000 | Disorders of menstruation and other abnormal bleeding from female genital tract | Female reproductive related |
| 628 | 5000 | Infertility, female | |
| 768 | 164 | Intrauterine hypoxia and birth asphyxia | Diseases of Newborn |
| 770 | 5000 | Other respiratory conditions of fetus and newborn | |

### 2.2 Cohort selection

We construct various sensitive disease patient cohorts and the non-sensitive disease patient cohort (consisting of patients without any diagnosis in the sensitive disease category) by querying Northwestern Medicine® Enterprise Data Warehouse (NMEDW). The NMEDW is a joint initiative across the Northwestern University Feinberg School of Medicine and Northwestern Memorial HealthCare [19]. NMEDW hosts a comprehensive and integrated repository of all clinical and research data sources from two hospitals: Northwestern Memorial Hospital (NMH) and Lake Forest Hospital (LFH). We use the following logic to construct the sensitive disease patient cohorts. For a particular ICD9 diagnosis code, we include all patients whose first diagnosis with the particular ICD9 code occurs between 10/1/2010 and 9/30/2015. We restrict the sample to active residents of NMH or LFH and randomly select 5,000 such patients if returned cohort is larger than 5,000 (see Table 1 for details). Most of the cohort sizes are on this order of magnitude. We kept the sample size roughly constant to eliminate the possible influence of the sample size on the classification results. When selecting the non-sensitive disease cohort, we note that many patients do not have any diagnoses codes associated with them (e.g., many healthy patients come to NMH or LFH for a physical examination). We thus further require that the patients must have at least one diagnosis code. For non-sensitive disease cohort, we randomly select 30,000 patients who visited NMH or LFH between 10/1/2010 and 9/30/2015 and never had a diagnosis listed as a sensitive disease. The patients with sensitive diseases often have many diagnoses, as shown in Figure 1 (a). Figure 1 (b) shows that even when we require the patients in the non-sensitive disease cohort to have at least one diagnosis, the distribution of diagnosis counts tends to skew towards the lower end compared to sensitive disease cohorts (e.g., Figure 1 (a)). The differences in the diagnosis counts between sensitive disease patient cohorts and non-sensitive disease cohort may introduce bias to the classification in the sense that a seemingly good or bad classifier in fact is influenced by the fact that some patients have more comorbid conditions while other patients have fewer comorbidities. To offset this bias, we conduct stratified sampling of patients from non-sensitive disease collection by matching the percentages of patients with a comparable number of diagnoses in the non-sensitive disease cohort to those from the unionized sensitive disease cohort.

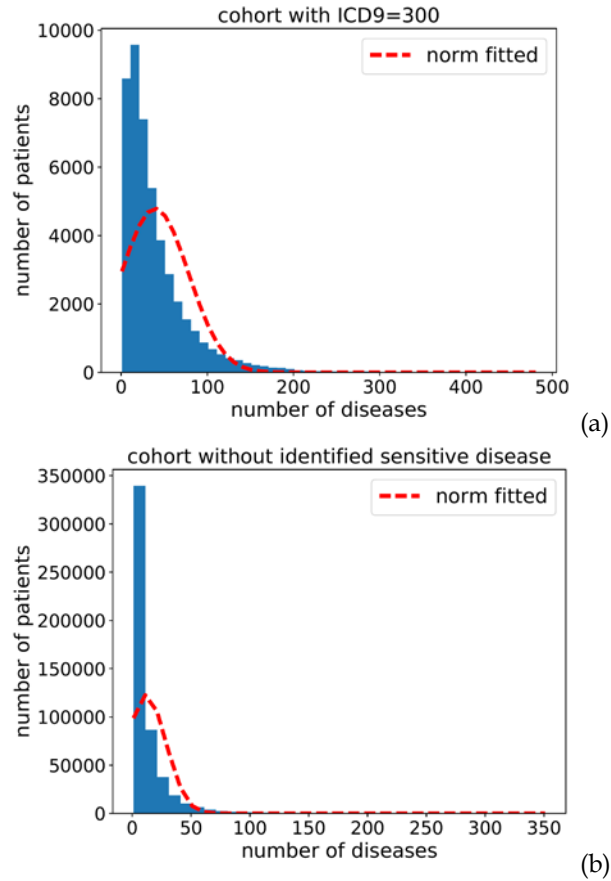

Figure 1 Distribution of number of diagnoses for patients in the cohort with an ICD9 code 300 (Anxiety, dissociative and somatoform disorders) and in the cohort without sensitive diseases. Both panels show distribution before random sampling.

## 2.3 Feature collection

We use the patients' laboratory test results, medications and procedures as their features. For a particular feature, we extract all measurements and records across a patient's entire medical record history. This is to support a worst-case analysis as there may be no limitation on what can be included in the dataset released for secondary analyses. We use binary encoding to represent the features. For laboratory test results, we compare their values with the corresponding low and high reference ranges presented in NMEDW. For example, if the Blood Urea Nitrogen (BUN) value is below low reference value, we set BUN_low to 1 and BUN_high to 0. With this binary encoding, the lab features should be interpreted as whether a patient has had an abnormally low or high value for a particular lab test. We apply similar binary encoding to medications, procedures, and non-sensitive diagnoses to capture whether a patient has taken a medication, has undergone a procedure, or has a comorbidity. The feature dimensions for the sensitive disease and non-sensitive disease cohorts are shown in Table 2.

Table 2 Number of features for sensitive disease and non-sensitive disease cohorts with matching diagnosis count (nonsd_match). Note that the total feature number for a particular cohort is 2×lab + procedure + med, because for each lab, there are 2 features: low and high.

| ICD9 | Medication | Procedure | Lab | Comorbidity |
|------|-----------|-----------|-----|-------------|
| 042 | 1768 | 2071 | 806 | 804 |
| 099 | 1219 | 1515 | 468 | 619 |
| 300 | 2274 | 3005 | 992 | 882 |
| 304 | 1843 | 2385 | 908 | 865 |
| 305 | 1994 | 2643 | 896 | 883 |
| 306 | 1790 | 2311 | 724 | 760 |
| 311 | 2339 | 3110 | 997 | 897 |
| 606 | 1338 | 1801 | 601 | 667 |
| 607 | 2172 | 2936 | 935 | 858 |
| 626 | 1998 | 2531 | 837 | 780 |
| 628 | 1782 | 2339 | 725 | 738 |
| 768 | 511 | 465 | 359 | 362 |
| 770 | 683 | 709 | 516 | 490 |
| nonsd_match | 2912 | 4099 | 1079 | 952 |

## 2.4 Dataset construction

When an EHR-based dataset is released for secondary analyses, the dataset typically contains a mixture of patients with sensitive diseases and with non-sensitive diseases. What immediately compromises privacy is a security attacker's ability to identify patients with a particular sensitive disease (e.g. Drug dependence with ICD9 code of 304). We approximate this scenario by using each sensitive disease (e.g. ICD9 code 304) as the case group, and leaving the rest of the sensitive diseases and the non-sensitive disease cohorts as the control group. We then perform a supervised classification on the case and control groups. We anticipate that, when trying to identify a particular sensitive disease (the case), attackers will have limited amount of training data (e.g., by annotating a few cases and control themselves). Thus, during cross validation, we perform stratified split of the entire dataset into the training and test datasets, according to a 1:9 ratio. Such a ratio maps to about 500 case- and 3000 + 500*12 control- patients being annotated, which is a significant annotation effort even for a deliberated attacker.

## 2.5 Classification

We use logistic regression with $l_2$ norm regularization to classify whether a patient belongs to a particular sensitive disease cohort. Logistic regression is chosen as the classification model due to its popularity in the machine learning and statistics community. Logistic regression aims to maximize the conditional likelihood of the class label given the features of a patient, as shown in equation ( 1 ). For an instance $x$ in equation ( 1 ), $f_i(x)$ denotes its $i$th feature, $w_{ic}$ is the weight associated with $i$th feature and $c$th class. For binary classification problem, we often set $w_{i0} = 0$, simplify the notation of $w_{i1}$ to $w_i$ and reduce the weights to a vector $\boldsymbol{w}$.

$$p(c|x) = \frac{\exp(\sum_{i=1}^{L} w_{ic} f_i(x))}{\sum_{c'=0}^{C} \exp(\sum_{i=1}^{L} w_{ic'} f_i(x))} \quad (1)$$

Maximizing the conditional probability for all training instances is equivalent to minimizing the summation of negative log-likelihood for all instances, as shown in equation ( 2 ):

$$-\sum_{j=1}^{N} \log \left( p(c^j | x^j) \right) + \lambda \|\boldsymbol{w}\|_2 \quad (2)$$

The term $\lambda \|\boldsymbol{w}\|_2$ in equation ( 2 ) is the regularization term, which is added in order to avoid overfitting. Here we use the $l_2$-norm of the weight vector $w$ as a gauge of model complexity. In this experiment, we choose $\lambda = 1$.

## 2.6 Feature ablation

Feature selection is a standard practice of tuning machine learning models. In this study, we try to identify important features that are most predictive for sensitive diseases, and remove them (hence using the term feature ablation). Univariate feature selection is a common choice for feature selection. Univariate feature selection typically works by computing a score that indicates the predicting power of a feature. In this study, we use the $\chi^2$ statistic between each feature and the class label as the feature scores for classification tasks, as shown in equation ( 3 ). Note that in our experiment, we adopt binary classification and binary features. In equation ( 3 ), $c$ denotes the index for class lable, $f$ denotes the feature value; $p_{cf}$ is the percentage of pa-tients with class $c$ and feature value $f$; $n$ is the total number of patients; $p_{c\cdot} = \frac{\sum_{f=0}^{1} p_{cf}}{n}$ and $p_{\cdot f} = \frac{\sum_{c=0}^{1} p_{cf}}{n}$ are the percentages of patients with class $c$ and feature value $f$, respectively:

$$\chi^2 = \sum_{c=0}^{1} \sum_{f=0}^{1} \frac{(p_{cf} - p_{c\cdot} p_{\cdot f})^2}{p_{c\cdot} p_{\cdot f} / n} \quad (3)$$

Intuitively, the higher the $\chi^2$ statistic, the more closely the the feature $f$ is related to the corresponding class label $c$. Thus the $\chi^2$ statistic can be used as the feature score.

Besides the $\chi^2$ statistic, other metrics can also be used as feature scores. The ANOVA F-test statistic is the ratio of between-group variability and with-in group variability. In the following formula, the numerator and denominator are the between-group variability and with-in group variability, respectively,

$$F = \frac{\sum_{c=0}^{1} n_c (\overline{f}_c - \overline{f})^2 / (2 - 1)}{\sum_{c=0}^{1} \sum_{j=1}^{n_c} (f_{cj} - \overline{f}_c)^2 / (n - 2)} \quad (4)$$

where $\overline{f}_c$ and $\overline{f}$ respectively denote the sample mean of the feature values in group $c$ and overall mean of the feature values for all patients respectively; $n_c$ is the number of patients in group $c$; $n$ is the total number of patients; $f_{cj}$ is the binary feature value for the $j$th patient in group $c$. Intuitively, a predictive feature can effectively separate the groups, so it should have a small with-in group variability and a large between-group variability, and hence a larger F-test statistic. Thus the higher the F statistic is, the more predictive the feature is. In this study, we select features based on the $k$ highest feature scores. For $k = 10, 20, ..., 50, 100, 200, ..., 1000$, we delete the $k$-best features in an iterative manner, and evaluate the classification performance on the remaining features.

## 2.7 Evaluation metrics

To evaluate the performance of our classification models and the impact of feature ablation, we plot the Receiver Operating Characteristic (ROC) curve and compute the Area Under the ROC curve (AUC), along with metrics including precision, recall, and F-measure. Let $TP$ be the number of true positives, $FP$ the number of false positives, and $FN$ the number of false negatives. We can use the following equations to calculate precision, recall, and F-measure:

$$Precision = \frac{TP}{TP + FP} \quad (5)$$

$$Recall = \frac{TP}{TP + FN} \quad (6)$$

$$F\text{-measure} = \frac{2 \cdot Precision \cdot Recall}{Precision + Recall} \quad (7)$$

## 2.8 Random Simulation

We hypothesize that in addition to lab tests, medications and procedures that are unique to sensitive diseases, the percentage differences between case and control on over-lapping features may also play a key role in the identifiability of sensitive diseases. To test this hypothesis, we perform two random simulation experiments. In each of the experiments, we randomly generate a binary feature matrix with the same size as the corresponding real-data matrix that combines the case and the control. Each row of the matrix corresponds to a patient, and each column corresponds to a feature. In the random feature matrix, feature values are randomly generated from a binomial distribution: the number of 1's (denoted as $k$) for a particular feature has the following probability $\binom{n}{k}p^k(1-p)^{n-k}$, where $n$ is the total number of patients and $p$ is the probability of 1 for the particular binary feature. The simulations aim to approximate the two extreme situations regarding the differences of parameter $p$. In the first simulation, for the $i$th feature, the case and the control have the same probability $p_i$, drawn from a uniform [0,1] distribution; while in the second simulation, for the $i$th feature, the case and the control have probabilities $p_i^{case}$ and $p_i^{ctrl}$, drawn independently from a uniform [0,1] distribution. We then follow the same classification model and configuration (e.g., train:test split) as described above. The ROC curves for the two simulations and the real-data classifications are shown in Figure 2. From the figure, we see that: when the majority of the features have different $p$ parameters for case and control, the sensitive disease patients can be perfectly identified; when each feature has the same $p$ parameters for case and control, the sensitive disease patients are non-identifiable. In reality, there are a small fraction of features with different $p$ parameters between case and control, and the bigger the fraction is, the more identifiable the sensitive disease patients are.

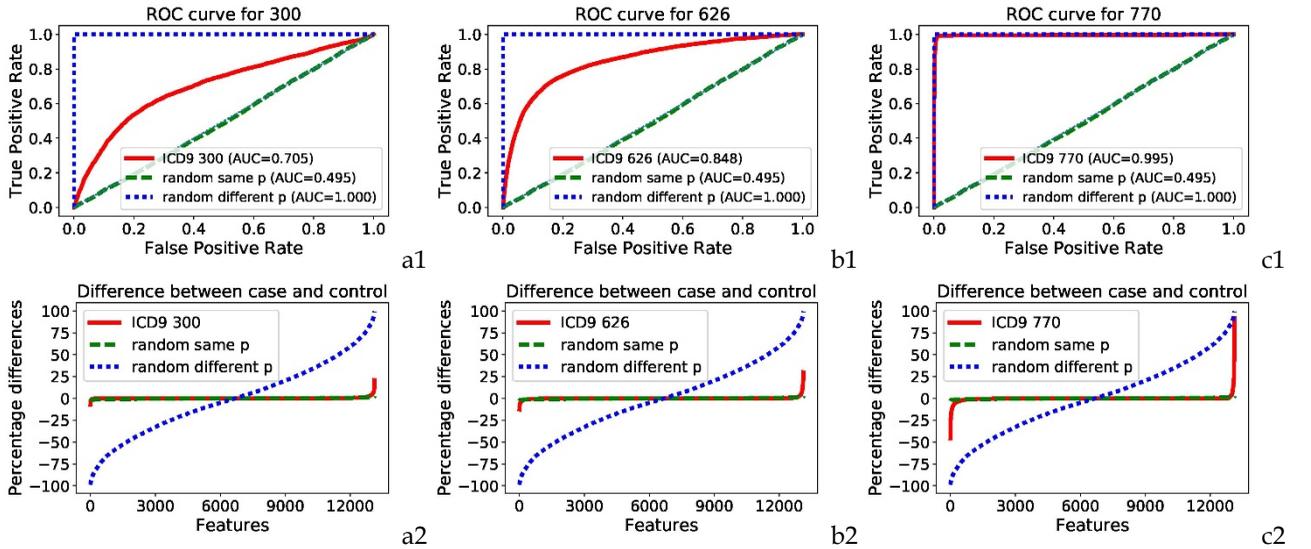

Figure 2 Random Simulations of identifiability of 3 typical sensitive diseases under different feature value distributions: ICD9 300 Anxiety, dissociative and somatoform disorders; ICD9 626 Disorders of menstruation and other abnormal bleeding from female genital tract; ICD9 770 Other respiratory conditions of fetus and newborn. Panels a1, b1, and c1 show ROC curves of the two random simulations and the real-data classification results using all features. Panels a2, b2, and c2 show the difference of percentages of 1's for each features ranked from low to high, for simulations and real data.

## 3 EXPERIMENTS AND RESULTS

As described in Section 2.3, our classification problem is formulated as distinguishing patients with a particular type of sensitive disease (case) from the patients without this sensitive disease (control). For a particular case control setting, we performed the following 10-fold cross validation. We equally split the data into 10 parts, with each part having the same number of case and control samples. We use each of one part for training, and the other 9 parts for testing, so that the train-test ratio is 1:9. We repeat 10 times to cover all of the case samples, and report the average of evaluation metrics.

When the non-sensitive disease cohort is selected by matching the distribution of diagnosis counts to those in the sensitive disease cohorts, the results of classification AUC for each sensitive disease as a function of feature ablations are shown in Table 3. The classification AUC curves can be grouped into 3 categories: (1) fast declining (AUC falls below 0.6 with less than 400 features excluded); (2) progressively declining (AUC falls below 0.7 with between 200 and 700 features excluded), (3) slowly declining (AUC does not fall below 0.7 with 1000 features excluded). 5 diseases (with ICD9 codes of 099, 300, 305, 306, 311) fall into the first category; 7 diseases (with ICD9 codes of 042, 304, 606, 607, 626, 628, 768) belong to the second category; and 1 disease (with ICD9 codes of 770) falls into the third category. The separation of 3 categories demonstrates the diversity of the sensitive diseases regarding the difficulty in identifying them. The majority of the diseases (12 out of 13) belong to the first two categories, suggesting that feature ablation is effective or has the potential of reducing sensitive disease identifiability.

Table 3 The impact of feature ablation on sensitive disease classification AUC with diagnosis-count-matching between sensitive and non-sensitive disaese cohort. The top row indicates whether using all features or excluding a certain number of top features. Red shades indicate AUC less than 0.7, whereas above 0.7 AUC is generally regarded as modest classification performance. Green shades indicate AUC less than 0.6, whereas 0.5 AUC means classification is no different than coin toss.

| ICD9 | All | 10 | 20 | 30 | 40 | 50 | 100 | 200 | 300 | 400 | 500 | 600 | 700 | 800 | 900 | 1000 |
|------|------|------|------|------|------|------|------|------|------|------|------|------|------|------|------|------|
| 099 | 0.783 | 0.692 | 0.674 | 0.664 | 0.658 | 0.656 | 0.645 | 0.627 | 0.614 | 0.599 | 0.592 | 0.581 | 0.576 | 0.565 | 0.560 | 0.555 |
| 300 | 0.676 | 0.606 | 0.591 | 0.585 | 0.582 | 0.576 | 0.565 | 0.553 | 0.542 | 0.536 | 0.530 | 0.523 | 0.518 | 0.515 | 0.512 | 0.507 |
| 305 | 0.698 | 0.660 | 0.653 | 0.647 | 0.646 | 0.642 | 0.634 | 0.614 | 0.604 | 0.593 | 0.583 | 0.572 | 0.560 | 0.551 | 0.548 | 0.543 |
| 306 | 0.739 | 0.640 | 0.619 | 0.608 | 0.599 | 0.593 | 0.574 | 0.558 | 0.547 | 0.540 | 0.535 | 0.534 | 0.526 | 0.521 | 0.520 | 0.515 |
| 311 | 0.711 | 0.628 | 0.609 | 0.604 | 0.604 | 0.599 | 0.589 | 0.572 | 0.560 | 0.553 | 0.547 | 0.539 | 0.533 | 0.526 | 0.521 | 0.517 |
| 042 | 0.914 | 0.862 | 0.836 | 0.824 | 0.815 | 0.802 | 0.760 | 0.722 | 0.697 | 0.680 | 0.658 | 0.643 | 0.635 | 0.628 | 0.617 | 0.617 |
| 304 | 0.866 | 0.834 | 0.822 | 0.815 | 0.812 | 0.810 | 0.785 | 0.762 | 0.744 | 0.727 | 0.709 | 0.694 | 0.670 | 0.657 | 0.646 | 0.639 |
| 606 | 0.910 | 0.834 | 0.773 | 0.755 | 0.747 | 0.747 | 0.734 | 0.716 | 0.699 | 0.686 | 0.673 | 0.660 | 0.642 | 0.631 | 0.626 | 0.612 |
| 607 | 0.851 | 0.771 | 0.759 | 0.748 | 0.743 | 0.737 | 0.720 | 0.691 | 0.668 | 0.645 | 0.630 | 0.619 | 0.600 | 0.589 | 0.582 | 0.573 |
| 626 | 0.826 | 0.809 | 0.796 | 0.788 | 0.783 | 0.780 | 0.758 | 0.730 | 0.694 | 0.672 | 0.658 | 0.648 | 0.634 | 0.618 | 0.609 | 0.599 |
| 628 | 0.910 | 0.878 | 0.875 | 0.874 | 0.864 | 0.849 | 0.832 | 0.792 | 0.768 | 0.747 | 0.716 | 0.693 | 0.679 | 0.664 | 0.651 | 0.639 |
| 768 | 0.834 | 0.833 | 0.832 | 0.835 | 0.834 | 0.832 | 0.840 | 0.840 | 0.818 | 0.779 | 0.756 | 0.725 | 0.698 | 0.668 | 0.642 | 0.620 |
| 770 | 0.993 | 0.992 | 0.991 | 0.991 | 0.991 | 0.991 | 0.990 | 0.986 | 0.980 | 0.973 | 0.966 | 0.957 | 0.952 | 0.939 | 0.928 | 0.917 |

Diseases with ICD9 codes 300 (Anxiety, dissociative and somatoform disorders), 626 (Disorders of menstruation and other abnormal bleeding from female genital tract), and 770 (Other respiratory conditions of fetus and newborn) are examples of sensitive diseases in each category, with decreasing difficulties (or increasing likelihood) to identify. The AUC of ICD9 300 is only 0.676 even if no feature is deleted, and quickly drops toward 0.6 after removing a few top features. The AUC of disease 626 is 0.826 without deleting any features, and quickly decreases below 0.8 after removing the top 20 features, and then drops below 0.7 after removing 300 features. Disease 770 starts with above 0.99 AUC and remains 0.9 after excluding 1000 top features.

Figure 3 shows the trends of precision, recall, F-measure, and AUC, under one random train:test split, as feature ablations progress for diseases with ICD9 code 300, 626, and 770, respectively. Three classification feature sets are used, including "union", "control", and "intersect". The "union" feature set includes features occurred in either the case or the control cohort. The "control" feature set includes features in the control cohort. The "intersect" feature set includes features occurred in both the case and the control cohort. Figure 3 also shows the number of features in different feature sets. Clearly, the number of features are roughly equal to the "union" and the "control" feature sets. Intersecting features results in significant drop in the number of features. It is not surprising to see that intersect feature sets correspond to the fastest classification performance decrease as there are (much) fewer features to begin with. Interestingly, diseases with fewer intersect features (when the case features overlap with fewer control features) tend to have slower performance degradation as feature ablation progresses, which suggests that in those cases, classifications largely rely on control features.

In general, for diseases in the first category (e.g., ICD9 300), they start with only moderate AUC. In addition, their precision, recall and F-measure start low (e.g., precision below 0.2, recall and F-measure below 0.1 for ICD9 300). For diseases in the second category (e.g., ICD9 626), they start with high AUC. Besides, their precision, recall and F-measure start with modest values (e.g., precision around 0.5, recall around 0.3, and F-measure around 0.4 for ICD9 626). But their precision, recall and F-measure quickly drop when we delete a small number of top features (e.g., 30 top features for ICD9 626). For diseases in the third category (e.g., ICD9 770), they start with over 0.99 AUC. Moreover, their precision, recall and F-measure start high and gradually decrease as feature ablation progresses. It is interesting to note that even when AUCs are high, the F-measure are only modest (e.g., 0.4 for ICD9 626). In fact, the first and second disease categories both seem to demonstrate reasonable non-identifiability for sensitive disease patients with moderate feature ablation efforts.

When the non-sensitive disease cohort is selected by requiring each patient has at least one diagnosis, the results of classification AUC for each sensitive disease as feature ablation progress are shown in Table 4. Compared to Table 3, Table 4 does not require diagnosis count matching hence is more relaxed. Table 4 demonstrates similar category structures of sensitive diseases regarding difficulty in identifying them. The values in Table 4 are similar to those in Table 3, suggesting that the classification performance is not very sensitive to whether the number of diagnoses is matched between sensitive and non-sensitive disease cohorts. Such insensitivity will likely save the effort for matching the severity between the patients with and without sensitive diseases (such effort may artificially alter patient distribution and introduce bias) when releasing the dataset for secondary analysis and preserving the non-identifiability of sensitive disease patients. An additional observation regarding both settings is that when ablating 1,000 out of a total of over 13,000 features (<8%), the predictability of most sensitive diseases is reduced (to <0.7 AUC) and many are close to poorly predictable (to <0.6 AUC).

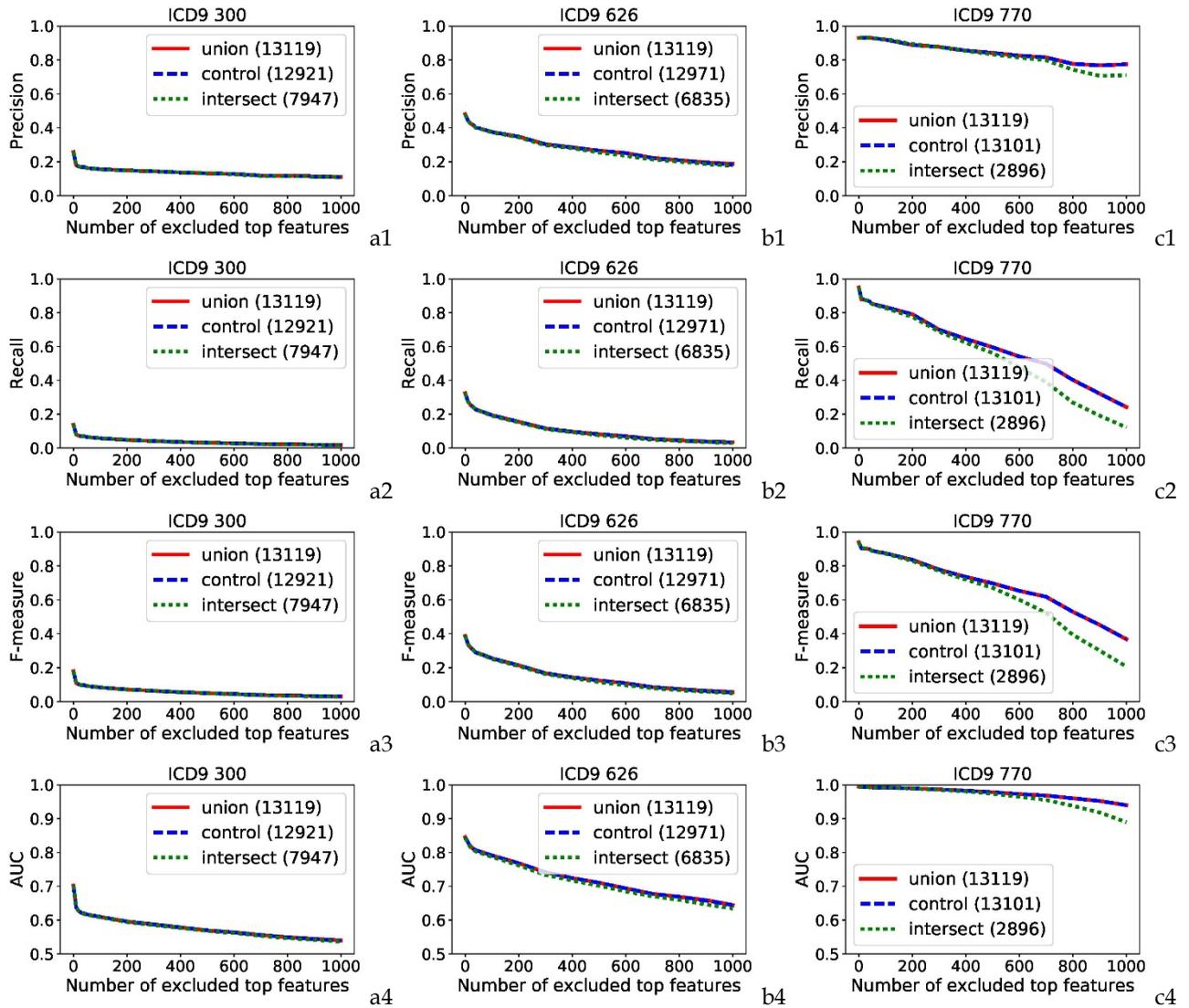

Figure 3 Trends of evaluation metrics as feature ablation progresses for three representative sensitive diseases: ICD9 300 Anxiety, dissociative and somatoform disorders; ICD9 626 Disorders of menstruation and other abnormal bleeding from female genital tract; ICD9 770 Other respiratory conditions of fetus and newborn. The panels a1 through a4 correspond to precision, recall, F-measure and AUC for ICD9 code 300, b1 through b4 correspond to ICD9 code 626, and c1 through c4 correspond to ICD9 code 770, respectively. Three classification feature sets are used, including "union", "control", and "intersect". The "union" feature set includes features occurred in either the case or the control cohort. The "control" feature set includes only features in the control cohort. The "intersect" feature set includes features occurred in both the case and the control cohort. Also shown in the legend of the figures are the number of features.

Table 4 The impact of feature ablation on sensitive disease classification AUC without diagnosis-count-matching between sensitive and non-sensitive disease cohort. Each patient in the non-sensitive disease cohort contains at least one diagnosis. The top row indicates whether using all features or excluding a certain number of top features. Red shades indicate AUC less than 0.7, whereas above 0.7 AUC is generally regarded as modest classification performance. Green shades indicate AUC less than 0.6, 0.5 AUC means classification is no different than coin toss.

| ICD9 | All | 10 | 20 | 30 | 40 | 50 | 100 | 200 | 300 | 400 | 500 | 600 | 700 | 800 | 900 | 1000 |
|------|-----|----|----|----|----|----|-----|-----|-----|-----|-----|-----|-----|-----|-----|------|
| 099 | 0.779 | 0.696 | 0.655 | 0.641 | 0.635 | 0.631 | 0.618 | 0.617 | 0.601 | 0.592 | 0.579 | 0.569 | 0.560 | 0.555 | 0.549 | 0.536 |
| 300 | 0.671 | 0.606 | 0.589 | 0.585 | 0.579 | 0.577 | 0.565 | 0.557 | 0.552 | 0.545 | 0.537 | 0.531 | 0.527 | 0.522 | 0.519 | 0.514 |
| 305 | 0.683 | 0.646 | 0.639 | 0.636 | 0.634 | 0.630 | 0.620 | 0.609 | 0.593 | 0.581 | 0.571 | 0.560 | 0.551 | 0.544 | 0.537 | 0.529 |
| 306 | 0.748 | 0.645 | 0.626 | 0.616 | 0.605 | 0.602 | 0.582 | 0.576 | 0.565 | 0.554 | 0.546 | 0.538 | 0.535 | 0.532 | 0.526 | 0.521 |
| 311 | 0.706 | 0.633 | 0.609 | 0.605 | 0.601 | 0.597 | 0.585 | 0.581 | 0.578 | 0.569 | 0.560 | 0.552 | 0.547 | 0.538 | 0.531 | 0.526 |
| 042 | 0.910 | 0.858 | 0.829 | 0.817 | 0.809 | 0.787 | 0.750 | 0.723 | 0.706 | 0.682 | 0.658 | 0.644 | 0.623 | 0.611 | 0.596 | 0.591 |
| 304 | 0.857 | 0.824 | 0.809 | 0.804 | 0.793 | 0.791 | 0.768 | 0.754 | 0.744 | 0.722 | 0.705 | 0.691 | 0.664 | 0.649 | 0.634 | 0.622 |
| 606 | 0.915 | 0.835 | 0.764 | 0.744 | 0.735 | 0.733 | 0.716 | 0.706 | 0.698 | 0.680 | 0.652 | 0.641 | 0.625 | 0.620 | 0.611 | 0.597 |
| 607 | 0.858 | 0.785 | 0.773 | 0.762 | 0.759 | 0.756 | 0.737 | 0.725 | 0.719 | 0.687 | 0.662 | 0.650 | 0.635 | 0.619 | 0.609 | 0.597 |
| 626 | 0.817 | 0.801 | 0.782 | 0.776 | 0.767 | 0.759 | 0.741 | 0.718 | 0.704 | 0.672 | 0.652 | 0.640 | 0.627 | 0.604 | 0.593 | 0.582 |
| 628 | 0.910 | 0.874 | 0.871 | 0.870 | 0.858 | 0.844 | 0.826 | 0.808 | 0.786 | 0.739 | 0.713 | 0.696 | 0.682 | 0.666 | 0.648 | 0.636 |
| 768 | 0.802 | 0.801 | 0.800 | 0.801 | 0.798 | 0.799 | 0.800 | 0.797 | 0.792 | 0.781 | 0.747 | 0.706 | 0.668 | 0.643 | 0.618 | 0.596 |
| 770 | 0.992 | 0.988 | 0.988 | 0.988 | 0.988 | 0.988 | 0.986 | 0.985 | 0.980 | 0.973 | 0.964 | 0.953 | 0.943 | 0.929 | 0.915 | 0.894 |

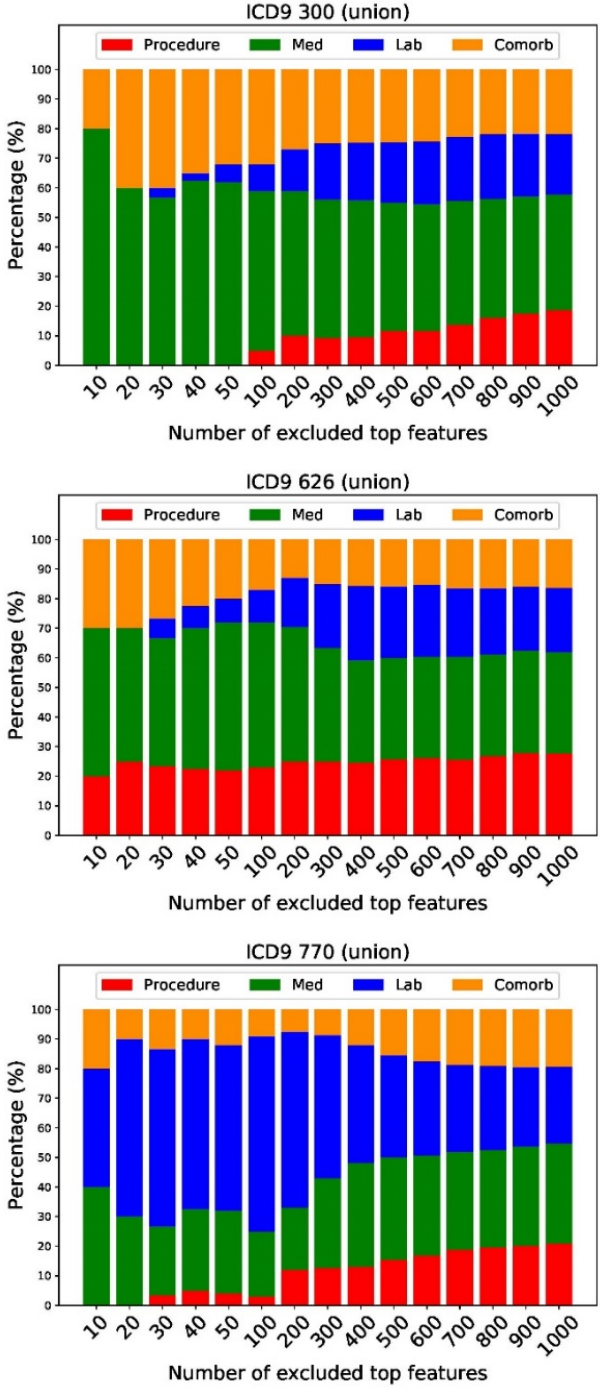

Figure 4 Top feature category percentage as feature ablation progresses.

Figure 4 shows the category (lab, medication, procedure features) percentage of excluded features as feature ablation progresses with increasing number of features. For most sensitive diseases in the first category, medication often rank as top features. For most sensitive diseases with slowly declining AUCs, lab features often play an important role. For sensitive diseases in the second category, the situation is more complex, each category of features may dominate the classification performance (e.g., procedures dominate top features for ICD9 626). Moreover, about 20% of ablated features are co-morbidity at the ablation level of 1000. Co-morbidity ablation seems to be the less compared to lab, procedure and medication features, across the three categories of ablation trends. Admittedly, losing comorbidities will make individual trajectories incomplete. However, an incomplete patients records might still have significant usefulness to researchers. For example, for the retained comorbidities which account of majority of comorbidities, their co-occurrence patterns and statistics will still be preserved. Hence the sensitive information ablated patients records should still be valuable to research targeting non-sensitive comorbidities.

Figure 5 shows $\chi^2$ feature scores and ANOVA F-test feature scores for the representative sensitive diseases. It is consistent with the intuition that sensitive diseases with faster decreasing AUCs have lower feature scores for top features. For example, ICD9 300 begins with $\chi^2$ feature scores less than $10^4$ and ANOVA F-test feature scores but ICD9 626 and 770 begin with feature scores more than $10^4$. In addition, sensitive diseases with slower decreasing AUCs tend to have slower decreasing feature scores. For example, at round $1000^{th}$ feature for ICD9 626, the $\chi^2$ feature scores and ANOVA F-test feature scores drop to 10; while at around $2000^{th}$ feature for ICD9 770, the $\chi^2$ feature scores and ANOVA F-test feature scores drop to 10. Also, to note that, the features in Figure 5 are ranked in descending order of $\chi^2$ feature scores in order to visually illustrate the different rankings of $\chi^2$ feature scores and ANOVA F-test feature scores. In ANOVA F-test feature scores, there are only slight jitterings of the scores, demonstrating that $\chi^2$ feature scores and ANOVA F-test feature scores largely agree with each other. It is for this reason that we only show the feature ablation results with ablation order based on $\chi^2$ feature scores in Table 3, Table 4, Figure 3 and Figure 4, because the feature ablation results with ablation order based on ANOVA F-test feature scores are very similar.

Table 5 shows the top features for classifying the three representative diseases according to $\chi^2$ feature scores. For ICD9 300 (Anxiety, dissociative and somatoform disorders), the top features are popoulated by anti-anxiety or anti-depression medications such as alprazolam, clonazepam, escitalopram etc; for ICD9 626 (Disorders of menstruation and other abnormal bleeding from female genital tract), the top features include female reproductive-system-related procedures, medications, and cormobidities. Finally, for ICD9 770 (Other respiratory conditions of fetus and newborn), we see labs and procedures specifically for neonates as well as medications such as Zinc Oxide which can treat and prevent diaper rash. In general, many of these features are uniquely associated with the specific sensitive diseases, which partly explains the higher feature scores for the top few features as shown in Figure 5.

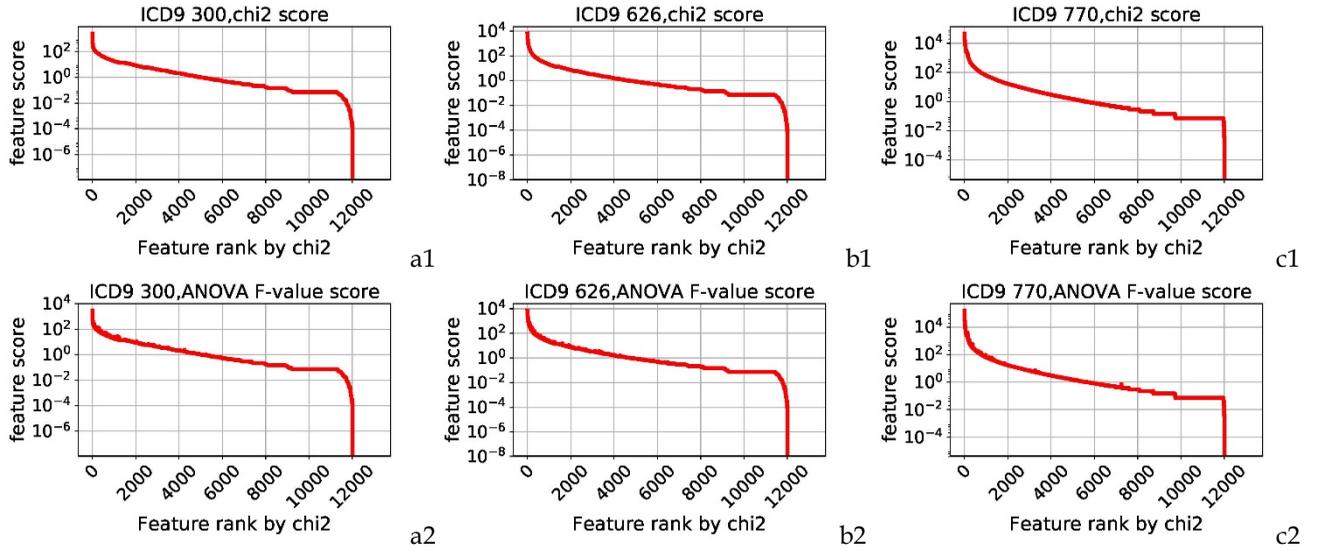

Figure 5 Feature scores for representative sensitive diseases: ICD9 300 Anxiety, dissociative and somatoform disorders; ICD9 626 Disorders of menstruation and other abnormal bleeding from female genital tract; ICD9 770 Other respiratory conditions of fetus and newborn. Panels a1, b1, c1 show the $\chi^2$ feature scores, while panels a2, b2, c2 show the ANOVA F-test feature scores. The features are ranked according to descending order of $\chi^2$ feature scores in order to visually illustrate the different rankings of $\chi^2$ feature scores and ANOVA F-test feature scores.

Table 5 Top 10 features for representative sensitive diseases: ICD9 300 Anxiety, dissociative and somatoform disorders; ICD9 626 Disorders of menstruation and other abnormal bleeding from female genital tract; ICD9 770 Other respiratory conditions of fetus and newborn.

| ICD9 300 | ICD9 626 | ICD9 770 |
|---|---|---|
| Med_alprazolam | Comor_218 | Comor_V30 |
| Med_escitalopram | Procedure_us pelvis complete | Lab_low_bedside glucose |
| Med_clonazepam | Comor_625 | Med_vitamin k 1 |
| Med_lorazepam | Med_bupivacaine-fentanyl | Med_erythromycin |
| Med_sertraline | Procedure_obstetrical ultrasound | Med_ampicillin |
| Med_citalopram | Med_oxytocin | Comor_V29 |
| Comor_780 | Med_lanolin | Lab_high_bilirubin-neonatal |
| Med_bupropion | Med_glycerin-witch hazel | Lab_high_bilirubin neonatal |
| Comor_V70 | Med_benzocaine | Med_zinc oxide |
| Med_fluoxetine | Comor_V25 | Lab_low_glucose |

## 4 Discussion and Future Work

Our results show that patients with different sensitive diseases have different identifiability and therefore their protection deserves customized efforts. Event-level privacy protection for some of the sensitive diseases requires removing a few related features but others require ablating many more. This study provides evidence to develop customized event-level privacy-preserving mechanisms and offer fine-grained protection. Comparing to the traditional de-identification method, ours provides an extra layer of protection by redacting events associated with the sensitive diseases so that even linkage attack for re-identification [20, 21] cannot reveal the diseases of concern to individual patients.

Our study also comes with several limitations. In our study, we selected the relatively prevalent sensitive diseases that produce more than 5000 patients during the retrospective time window between 10/1/2010 and 9/30/2015. However, there are other diseases that are sensitive but less prevalent. The classification tasks for identifying less prevalent sensitive diseases may present unique challenges in both model selection and feature engineering. For example, when there are only a few positive labels, the classification tasks may be formulated as an anomaly detection problem. In addition, although the evaluation is based on the data from two hospitals, they are from the same region (Chicago area) and may present only limited geographic diversity. We plan to test our findings from multiple institutions in research network setting [22] at national scale and check the consistency of the results.

In this study, we formulated the features as whether or not the patient has experienced a high/low lab test result, whether the patient has taken a certain medication, and whether a certain procedure has been performed on the patient. Feature ablation under this formulation will eliminate all occurrences of a particular lab test being high/low, all history of a particular medication, or all history of a particular procedure. One can choose to keep the timestamps of the features and increase the granularity of the event in order to preserve more medical events that occurred at other times and are not predictive of the sensitive disease a patient may have. In reality, attackers may also employ imputation methods and potentially obtain more abnormal lab test results, with and without timestamps [23, 24]. From this perspective, the fine-grained event-level privacy protection with attackers having imputation capability presents more challenges. However, even in finer grained event-level privacy protection, the formulation and results of our study cannot be bypassed because attackers can always reduce the problem to our formulation in order to have less sparse features and potentially better identifiability. We plan to systematically study such trade-offs as a next step.

This study provides a proof-of-concept feature ablation framework towards event-level privacy protection. We envision the integration of such fine-grained protection mechanism with informed consent to patients would encourage data sharing without compromising their privacy. We chose logistic regression with $l_2$-norm regularization as the classification model due to its popularity in both the machine learning and statistics community. Our feature ablation framework is general and applies to other classification models such as support vector machines, artifial neural networks, decision trees, random forests and Bayesian networks. One can also choose from other regularizations such as $l_1$-norm regularization. On the other hand, attackers may employ feature dimensionality reduction techniques when feature matrix is sparse [25, 26], or choose to model different categories of features using a tensor [27-31] to explore feature interaction, or use deep learning methods [32, 33], in order to improve the classification performance. The exact performances by using other classification models, regularizations, feature formulation and learning, are expected to vary depending on the combinations to be used. Future studies are encouraged to explore the above combinations of alternatives and better characterize the feasibility and the efficacy for the broader territory of event-level privacy protection. We have not integrated event-level privacy protection in a daily clinical workflow and will investigate the pipeline of this method at work in a real-world scenario in the future.

## 5 Conclusion

We conduct a study of event-level privacy protection, which might have the potential to provide customized protection of individuals' data. We demonstrated that patients with different sensitive diseases have different identifiability and therefore their protection deserves different mechanisms. We developed a new and intuitive feature ablation method to protect event-level privacy. From a group of sensitive diagnoses, we evaluated the feasibility and the efficacy of our proposed method. As feature ablation progresses, the AUCs decrease with varying speeds, for classifying patients with sensitive disease vs. other patients. The sensitive diseases can be divided into 3 categories according to the trends that feature ablation decreases their identifiability: fast declining (AUC falls below 0.6 with less than 400 features excluded), progressively declining (AUC falls below 0.7 with between 200 and 1000 features excluded), slowly declining (AUC does not fall below 0.7 with 1000 features excluded). The fact that the majority (12 out of 13) of the sensitive diseases fall into the first two categories suggests the practical applicability or potential of feature ablation for event-level privacy protection.

## 6 Acknowledgment

This work was supported in part by NIH Grant 1R21LM012618-01. Drs. Chengsheng Mao and Yuan Zhao are co-first authors. Dr. Yuan Luo is the corresponding author.

## 7 References

[1]   *Health Insurance Portability and Accountability Act (HIPAA)*. Available: http://www.hhs.gov/ocr/hipaa

[2]   P. Kwok, M. Davern, E. Hair, and D. Lafky, "Harder than you think: a case study of re-identification risk of HIPAA-compliant records," *Chicago: NORC at The University of Chicago. Abstract,* vol. 302255, 2011.

[3]   K. El Emam, E. Jonker, L. Arbuckle, and B. Malin, "A systematic review of re-identification attacks on health data," *PloS one,* vol. 6, no. 12, p. e28071, 2011.

[4]   Ö. Uzuner, Y. Luo, and P. Szolovits, "Evaluating the state-of-the-art in automatic de-identification," *Journal of the American Medical Informatics Association,* vol. 14, no. 5, pp. 550-563, 2007.

[5]   J. Vaidya, B. Shafiq, X. Jiang, and L. Ohno-Machado, "Identifying inference attacks against healthcare data repositories," *AMIA Jt Summits Transl Sci Proc,* vol. 2013, pp. 262-6, 2013.

[6]   M. Gymrek, A. L. McGuire, D. Golan, E. Halperin, and Y. Erlich, "Identifying personal genomes by surname inference," *Science,* vol. 339, no. 6117, pp. 321-4, Jan 18 2013.

[7]   A. Stubbs, C. Kotfila, and Ö. Uzuner, "Automated systems for the de-identification of longitudinal clinical narratives: Overview of 2014 i2b2/UTHealth shared task Track 1," *Journal of biomedical informatics,* vol. 58, pp. S11-S19, 2015.

[8]   S. Xu, S. Su, L. Xiong, X. Cheng, and K. Xiao, "Differentially private frequent subgraph mining," in *Data Engineering (ICDE), 2016 IEEE 32nd International Conference on,* 2016, pp. 229-240: IEEE.

[9]   C. Dwork, "Differential privacy," in *Encyclopedia of Cryptography and Security*: Springer, 2011, pp. 338-340.

[10]  X. Jiang, Z. Ji, S. Wang, N. Mohammed, S. Cheng, and L. Ohno-Machado, "Differential-private data publishing through component analysis," *Transactions on data privacy,* vol. 6, no. 1, p. 19, 2013.

[11] N. Mohammed, X. Q. Jiang, R. Chen, B. C. M. Fung, and L. Ohno-Machado, "Privacy-preserving heterogeneous health data sharing," (in English), *Journal of the American Medical Informatics Association,* vol. 20, no. 3, pp. 462-469, May 2013.

[12] H. R. Li, L. Xiong, L. Ohno-Machado, and X. Q. Jiang, "Privacy Preserving RBF Kernel Support Vector Machine," (in English), *Biomed Research International,* 2014.

[13] H. Li, L. Xiong, L. Zhang, and X. Jiang, "DPSynthesizer: Differentially Private Data Synthesizer for Privacy Preserving Data Sharing," *Proceedings VLDB Endowment,* vol. 7, no. 13, pp. 1677-1680, Aug 2014.

[14] H. Li, L. Xiong, and X. Jiang, "Differentially Private Histogram and Synthetic Data Publication," in *Medical Data Privacy Handbook*: Springer, 2015, pp. 35-58.

[15] H. Li, L. Xiong, Z. Ji, and X. Jiang, "Partitioning-Based Mechanisms Under Personalized Differential Privacy," in *Pacific-Asia Conference on Knowledge Discovery and Data Mining*, 2017, pp. 615-627: Springer.

[16] G. Hripcsak and D. J. Albers, "Next-generation phenotyping of electronic health records," *Journal of the American Medical Informatics Association,* vol. 20, no. 1, pp. 117-121, 2012.

[17] C. Shivade *et al.*, "A review of approaches to identifying patient phenotype cohorts using electronic health records," *Journal of the American Medical Informatics Association,* vol. 21, no. 2, pp. 221-230, 2013.

[18] P. Raghavan, J. L. Chen, E. Fosler-Lussier, and A. M. Lai, "How essential are unstructured clinical narratives and information fusion to clinical trial recruitment?," *AMIA Summits on Translational Science Proceedings,* vol. 2014, p. 218, 2014.

[19] J. B. Starren, A. Q. Winter, and D. M. Lloyd - Jones, "Enabling a learning health system through a unified enterprise data warehouse: the experience of the Northwestern University Clinical and Translational Sciences (NUCATS) Institute," *Clinical and translational science,* vol. 8, no. 4, pp. 269-271, 2015.

[20] L. Sweeney, "Uniqueness of simple demographics in the US population," in "Technical report," Carnegie Mellon University2000.

[21] P. Golle, "Revisiting the uniqueness of simple demographics in the US population," in *Proceedings of the 5th ACM workshop on Privacy in electronic society*, 2006, pp. 77-80: ACM.

[22] J. Yuan *et al.*, "Towards a privacy preserving cohort discovery framework for clinical research networks," *Journal of biomedical informatics,* vol. 66, pp. 42-51, 2017.

[23] Y. Luo, P. Szolovits, A. S. Dighe, and J. M. Baron, "3D-MICE: integration of cross-sectional and longitudinal imputation for multi-analyte longitudinal clinical data," (in eng), *J Am Med Inform Assoc,* Nov 30 2017.

[24] Y. Luo, P. Szolovits, A. S. Dighe, and J. M. Baron, "Using Machine Learning to Predict Laboratory Test Results," *American Journal of Clinical Pathology,* vol. 145, no. 6, pp. 778-788, 2016.

[25] F. Wang, P. Zhang, B. Qian, X. Wang, and I. Davidson, "Clinical risk prediction with multilinear sparse logistic regression," in *Proceedings of the 20th ACM SIGKDD international conference on Knowledge discovery and data mining*, 2014, pp. 145-154: ACM.

[26] Y. Luo, Y. Xin, R. Joshi, L. Celi, and P. Szolovits, "Predicting ICU Mortality Risk by Grouping Temporal Trends from a Multivariate Panel of Physiologic Measurements," in *Proceedings of the 30th AAAI Conference on Artificial Intelligence*, 2016.

[27] Y. Luo, F. Wang, and P. Szolovits, "Tensor factorization toward precision medicine," *Briefings in Bioinformatics,* March 19, 2016 2016.

[28] J. C. Ho *et al.*, "Limestone: High-throughput candidate phenotype generation via tensor factorization," *Journal of biomedical informatics,* vol. 52, pp. 199-211, 2014.

[29] J. C. Ho, J. Ghosh, and J. Sun, "Marble: high-throughput phenotyping from electronic health records via sparse nonnegative tensor factorization," in *Proceedings of the 20th ACM SIGKDD international conference on Knowledge discovery and data mining*, 2014, pp. 115-124: ACM.

[30] Y. Wang *et al.*, "Rubik: Knowledge guided tensor factorization and completion for health data analytics," in *Proceedings of the 21th ACM SIGKDD International Conference on Knowledge Discovery and Data Mining*, 2015, pp. 1265-1274: ACM.

[31] Y. Luo, F. S. Ahmad, and S. J. Shah, "Tensor factorization for precision medicine in heart failure with preserved ejection fraction," *Journal of Cardiovascular Translational Research,* pp. 1-8, 2017.

[32] R. Miotto, F. Wang, S. Wang, X. Jiang, and J. T. Dudley, "Deep learning for healthcare: review, opportunities and challenges," (in eng), *Brief Bioinform,* May 06 2017.

[33] R. Miotto, L. Li, B. A. Kidd, and J. T. Dudley, "Deep patient: An unsupervised representation to predict the future of patients from the electronic health records," *Scientific reports,* vol. 6, 2016.

**Chengsheng Mao** received the BS degree in Computer Science and Technology from Huazhong University of Science and Technology in 2008, and the PhD degree in Computer Application Technology from Lanzhou University in 2016. He is now a postdoctoral fellow in the Department of Preventive Medicine at Feinberg School of Medicine, Northwestern University. His research interests include data mining, machine learning in a range of domains including bioinformatics and electric healthcare.

**Yuan Zhao** received B.S. and M.S. degrees in telecommunications engineering from Beijing University of Posts and Telecommunications in 2009 and 2012, respectively, and PhD degree in electrical engineering from Arizona State University in 2016. He joined the Department of Preventive Medicine at the Northwestern University Feinberg School of Medicine as a postdoctoral fellow in 2017. He is recipient of the best student paper award of 43rd IEEE photovoltaic specialist conference in 2016. His current research interests include natural language processing of medical data and machine learning for computational phenotyping.

**Mengxin (Ivy) Sun**, MPP 2016, Bachelor of Law 2014, is currently an Analytics Consultant working at Northwestern Medicine, where she provided analytics support for various research projects. Her current research interests include using EMR data to inform clinical decision making and clinical operation optimization. And she has published on Journal of Hospital Medicine.

**Yuan Luo** is an assistant professor in the Department of Preventive Medicine at Northwestern University Feinberg School of Medicine. He received his PhD in Computer Science from Massachusetts Institute of Technology. He served on the student editorial board of Journal of American Medical Informatics Association. His research interests include machine learning, natural language processing, time series analysis, computational genomics, with a focus on biomedical applications. Dr. Luo is the recipient of the inaugural Doctoral Dissertation Award Honorable Mention by American Medical Informatics Association (AMIA) in 2017.